\documentclass{elsarticle}
 \usepackage{amssymb}
 \usepackage{amsfonts,amsmath,epsfig,psfrag}
 \usepackage{graphicx}
 \usepackage{palatino}
 \usepackage{natbib}
\bibpunct{[}{]}{;}{n}{}{,}
\def\rr#1{(\ref{#1})}

\renewcommand{\theequation}{\thesection.\arabic{equation}}

\def\pafrac#1#2{\frac{\partial #1}{\partial #2}}
\def\dd#1#2{\frac{d #1}{d #2}}

\def\boldsigma{\mbox{\boldmath $\sigma$}}
\def\boldkappa{\mbox{\boldmath $\kappa$}}

\begin{document}

\begin{frontmatter}
\title{Axisymmetric Indentation of Curved Elastic Membranes by a Convex Rigid Indenter}

\cortext[cor1]{Corresponding author, now at University of Manchester}
\address[pcs]{Department of Plant and Crop Sciences, School of Biosciences, University of Nottingham, Nottingham, LE12 5RD, UK}
\address[sms]{School of Mathematical Sciences, University of Nottingham, Nottingham, NG7 2RD, UK}

\author[pcs]{S.P.Pearce\corref{cor1}}
\ead{simon.pearce@manchester.ac.uk (current)}
\author[sms]{J.R.King}
\ead{john.king@nottingham.ac.uk}
\author[pcs]{M.J.Holdsworth}
\ead{michael.holdsworth@nottingham.ac.uk}

\begin{abstract}
Motivated by applications to seed germination, we consider the transverse deflection that results from the axisymmetric indentation of an elastic membrane by a rigid body. The elastic membrane is fixed around its boundary, with or without an initial pre-stretch, and may be initially curved prior to indentation. General indenter shapes are considered, and the load-indentation curves that result for a range of spheroidal tips are obtained for both flat and curved membranes. Wrinkling may occur when the membrane is initially curved, and a relaxed strain-energy function is used to calculate the deformed profile in this case. Applications to experiments designed to measure the mechanical properties of seed endosperms  are discussed.
\end{abstract}


\end{frontmatter}

\section{Introduction}
An elastic shell may be defined as a three-dimensional elastic body with one dimension that is much thinner than the other two \citep{libai1998}. This enables us to consider the deformation of only the mid-plane of the elastic body in this thinner direction. A membrane may then be defined as a shell which has negligible resistance to bending \citep{libai1998, humphrey2003}.  Elastic membranes are commonly found in biological and engineering contexts, where they may span relatively large areas despite having little volume or weight \citep{humphrey1998}. 

There are two different classes of non-linear elastic membrane theories and we refer to \citet{haughton2001} for a more complete comparison of the two theories, but give a brief description here. The first of these membrane theories may be called a \lq membrane-like shell\rq \citep{libai1998}, which takes the shell theory of three-dimensional elasticity and introduces the membrane assumption of no stress in the direction normal to the membrane. In this formulation the thickness is included in the derivation, as the principal stretch in the thickness direction, $\lambda_n$, appears in the governing equations, which explicitly allows the membrane to get thinner to conserve mass when the constraint of incompressibility is imposed. This is a common method of treating membranes, and a comprehensive derivation may be found in \citet{libai1998} and \citet{steigmann2007}. A variational treatment may also be considered, for example see \citet{ledret1996}.

The second class of membrane theories occurs from considering the mid-plane of the membrane as a two-dimensional sheet of elastic material embedded in  three-dimensional space, entirely neglecting thickness effects through the membrane \citep{li1995}. \citet{nadler2009} call this a \lq simple membrane', and a consequence of this reduction is that the stretch through the membrane is not included in the formulation. The deformation gradient is then two-dimensional, and there are only two strain invariants and two principal stretches, rather than the three that arise in the three-dimensional theory,  see \citet{steigmann2001,steigmann2009} for details.  

It has been shown that the simple membrane and membrane-like shell approaches give the same governing equations to leading order \citep{naghdi1972, haughton2001, steigmann2007}, and it is the specification of the constitutive behaviour which varies between the two theories. It is possible to define two-dimensional strain-energy functions which have no counterpart in the three-dimensional theory, as discussed in \citet{haughton2001}, as well as to use three-dimensional strain-energy functions in the two-dimensional theory. Further details of such matters may be found in \citet{libai1998} and  \citet{steigmann2007}.

When considering deformations of membranes using either of the above theories, it is important to ensure that the membrane is in a state of tension, with both principal stresses remaining positive throughout the deformation \citep{libai1998}. If compressive (negative) stresses occur then the membrane may wrinkle, a local buckling event \citep{miyamura2000} where alternating crests and troughs appear parallel to the direction of principal tension \citep{libai2002}, at which point additional refinements are required. In particular, \citet{andra2000} show that equilibrium states with distributions of infinitesimal wrinkles are possible when bending stiffness is neglected, due to loss of convexity of the strain-energy function. \citet{pipkin1986} showed how to modify the constitutive behaviour of the material in regions where compressive stresses exist, such that a pseudo-surface is considered which is in simple tension in these regions, neglecting the local details of the wrinkled structure in favour of global results; this is called tension field theory. For further details on treatments of wrinkling see for example, \citet{pipkin1986}, \citet{steigmann1990}, \citet{epstein2001ani} and \citet{libai2002}. A physical example of such wrinkles may be seen when a pointed object is pushed into a fixed curved piece of the plastic wrap, causing straight crests to radiate from the point of contact.

The indentation of an axisymmetric elastic membrane by a rigid body has been considered previously, although mostly through the use of linear elasticity. Here we are interested in large deformations, and therefore use non-linear elasticity theory to model the deflection of the membrane. Physical applications/occurrences of such indentations include puncture of rubber gloves by medical needles \citep{nguyen2009a, nguyen2009b}, stones embedding into rubber tyres \citep{yang1971} and our specific interest of seed germination, which we shall discuss later.

\citet{yang1971} considered the indentation of a flat circular hyperelastic membrane by a rigid sphere, in the large deformation regime, using the three-dimensional Mooney-Rivlin strain-energy function. \citet{liu2001} estimate the elastic modulus of a specific elastomeric film by comparing numerical simulations of the model introduced by \citet{yang1971} with their experimental observations. \citet{steigmann2005} and \citet{nadler2006} consider the indentation and penetration of a material described by a compressible two-dimensional Varga strain-energy function by a spherical-tipped cylindrical indenter, and assume that the membrane will break if a punctured state, with a discontinuous membrane, is energetically favourable. \citet{selvadurai2006} compares finite element simulations of indentation with a spherical indenter with experimental results, and provides a thorough literature review on rubber membranes, including a discussion of friction effects.

This rigid-body indentation is closely related to the adhesion behaviour of a rigid punch attached to a membrane, which exhibits stretching as the rigid body is pulled away. This has been considered by, for example, \citet{nadler2008}, who show that non-linear elastic effects are important; they also find that the imposition of pre-stress on the material has a strong effect on the force required to separate the punch from the membrane.

\citet{haughton1996} consider the response of planar and hemispherical elastic membranes to filling with a specific volume of incompressible liquid, using a relaxed strain-energy function when the membrane becomes wrinkled. They find an instability whereby the depth of the liquid increases very rapidly with only a slight increase in volume. Depending on the strain-energy function this unstable behaviour may continue indefinitely or re-stabilise, corresponding to the bifurcation points of a uniformly inflated spherical membrane. \citet{haughton1996} find that the plane membrane is a special case in that it remains tense throughout the deformation, whereas a hemispherical membrane contains a wrinkled section when the volume of liquid is small, which also reappears briefly after the instability.

In this article we shall generalise the previous work in three directions. Firstly, we allow the profile of the indenter tip to be of general axisymmetric shape rather than only considering spherical indenters, although we shall exclude sharply pointed tips by imposing mild continuity requirements. Secondly, we wish to investigate membranes which are initially curved, introducing this change of reference configuration into the elastic equilibrium equations. Lastly, we also wish to compare the effects of using different strain-energy functions to describe the constitutive behaviour of the membrane. 

Our primary motivation behind considering this deformation lies in seed germination, in particular the stretching and rupture of the endosperm by the root-tip in certain species of germinating seeds, specifically the two related Brassicaceae species \textit{Arabidopsis thaliana} \citep{holdsworth2008} and \textit{Lepidium sativum} \citep{muller2006}. These two species have approximately ellipsoidal seeds, including a thin, deformable tissue called the endosperm which surrounds the embryo. During the germination process the embryo grows, pushing the root-tip against the endosperm, stretching it before rupture occurs. The shape of the root-tip is approximately prolate spheroidal, as is the surrounding endosperm, hence our interest here in non-flat membranes and non-spherical indenter tips. %In addition it is important to consider strain-energy functions which are suitable to modelling biological materials.

The endosperm itself is thin and highly curved, and rapidly dehydrates when separated from the rest of the seed. It is therefore difficult to use conventional techniques to measure the elastic properties of this tissue, and so an indirect method is required. We are particularly interested in determining the elastic properties of the endosperm as biological changes occur during germination which alter the cell wall microstructure.

We therefore wish to lay the mathematical groundwork before modelling biological puncture force experiments in a later work. In these experiments the seed is dissected and the embryo removed before a metal indenter is pushed through the endosperm, examples of such experiments are given in \citet{muller2006}. The force and position of the indenter are measured during this experiment, and we then wish to estimate the elastic properties of the endosperm using this technique, particularly how these properties change over time. It has been shown in \citet{muller2006} that the force required to puncture through \textit{Lepidium sativum} endosperms decreases during the germination timescale. In preliminary experiments it has been shown that the vertical length of endosperm may double in size before rupture, clearly requiring non-linear elasticity to model this effect.

\section{Theory}
\subsection{Governing Equations}
We shall use the theory of non-linear elasticity as detailed in \citet{ogden1997} to model the deformation. We assume that the midsurface of the undeformed membrane has symmetry about a particular axis, and may therefore be described by cylindrical coordinates with respect to the usual basis $(\mathbf{e}_r,\mathbf{e}_\theta,\mathbf{e}_z)$, 
\begin{equation}\label{undefcon}
\mathbf{X} = \hat{R}(\hat{S}) \mathbf{e}_r(\Theta) + \hat{Z}(\hat{S}) \mathbf{e}_z, \; 0 \leq \hat{S} \leq \hat{L}, \; 0 \leq \Theta \leq 2 \pi, 
\end{equation}
where $\hat{S}$ is the arclength along the reference midsurface, which will be used as the independent variable throughout. We shall take the origin of our coordinates to be the centre of the membrane, $\hat{R}(0) = 0, \hat{Z}(0) = 0$, as well as requiring that the membrane be smooth there, $\hat{Z}'(0) = 0$.  If the membrane is initially flat we set $\hat{R}(\hat{S}) = \hat{S}, \hat{Z}(\hat{S}) = 0$, in this case $\hat{R}$ may be used as the independent coordinate as has been used in previous studies.

We are interested in axisymmetric deformations that map the reference configuration $\mathbf{X}$ to the deformed configuration $\mathbf{x}$, the midsurface of which is given by
\begin{equation}\label{defcon}
\mathbf{x} = \hat{r}(\hat{S}) \mathbf{e}_r(\theta) + \hat{z}(\hat{S}) \mathbf{e}_z, \; 0 \leq \hat{S} \leq \hat{L}, \; 0 \leq \theta \leq 2 \pi.
\end{equation}
A point in the deformed surface is therefore given by two variables, $\hat{r}$ and $\hat{z}$, the three-dimensional shape being then given by rotation around the $z$ axis. Immediately, due to the imposed axisymmetry, we may identify $\Theta = \theta$. We also impose the same conditions as in the undeformed state at the axis, that is $\hat{r}(0) = \hat{z}(0)=0, \hat{z}'(0)=0$.

Here we shall solve the static problem, where we assume that there is no time-dependence involved in the deformation. This is appropriate to the quasi-steady cases in which we are interested, for which the indentation is slow compared to the inertia of the membrane.

Appendix A outlines the derivation of the equilibrium equations for a curved  membrane, using the membrane-like shell approximation to the three-dimensional theory of elasticity, as discussed above. 
%From this starting point we may then use the two-dimensional theory of elastic surfaces to calculate the equilibrium equations, via calculating the tangent vectors and metric tensors involved, as detailed in \citet{steigmann2001}. In particular, using this theory gives an immediate base from which to study the indentation of an elastic shell rather than a membrane by including contributions to the strain-energy from bending as well as stretching.
%Alternatively, we may use the three-dimensional theory such as is used in \citet{yang1971} or \citet{haughton1996}. Both of these approaches give the same governing equations, subject to some conditions, although the difference in specification of the strain-energy function between the two is important. We choose here to use three-dimensional strain-energy functions, and the constraint of incompressibility. An outline of the derivation of the governing equations is given in Appendix A, although for the full derivation we refer to \citet{haughton2001}.
The principal stretch ratios may be defined in the circumferential, radial, and normal directions in the form (see Appendix A for details),
\begin{equation}\label{lambdas}
\lambda_s = \dd{\hat{s}}{\hat{S}} = \frac{\sqrt{\left(\dd{\hat{r}}{\hat{S}}\right)^2 + \left(\dd{\hat{z}}{\hat{S}}\right)^2}}{\sqrt{\left(\dd{\hat{R}}{\hat{S}}\right)^2 + \left(\dd{\hat{Z}}{\hat{S}}\right)^2}} , \quad \lambda_\theta = \frac{\hat{r}(\hat{S})}{\hat{R}(\hat{S})}, \quad \lambda_n = \frac{\hat{h}}{\hat{H}}, 
\end{equation}
where $\hat{h}$ and $\hat{H}$ are the deformed and undeformed thicknesses and $\hat{s}$ is the arclength in the deformed configuration. We assume that the membrane is made of a hyperelastic material with a strain-energy function given by  $\hat{W}(\lambda_s,\lambda_\theta,\lambda_s^{-1} \lambda_\theta^{-1})$, where we have already assumed that the material is incompressible and therefore set $\lambda_s \lambda_\theta \lambda_n = 1$; see \citet{ogden1997} for more details. The principal stress resultants per unit length in the deformed membrane are therefore defined as
\begin{equation}\label{stressres}
\hat{T}_s = \frac{\hat{H}}{\lambda_\theta} \pafrac{\hat{W}}{\lambda_s}, \quad
\hat{T}_\theta = \frac{\hat{H}}{\lambda_s} \pafrac{\hat{W}}{\lambda_\theta},
\end{equation}
which may be seen to be $\hat{T}_\alpha = \hat{h} \hat{\sigma}_\alpha$, where $\hat{\sigma}_\alpha = \lambda_\alpha \pafrac{\hat{W}}{\lambda_\alpha}$ is the usual principal Cauchy stress in incompressible three-dimensional elasticity, and similarly for $\hat{T}_\theta$. Appendix A shows how the stresses only occur in the combination $\hat{h}\hat{\sigma}_\alpha$, and hence we introduce $\hat{T}_\alpha$ with dimensions of force per length.

While the theory presented here is appropriate for any isotropic incompressible strain-energy function, we shall study some specific examples in order to make later comparisons. One commonly used such function is the Mooney-Rivlin strain-energy function,
\begin{equation}\label{mooney}
\hat{W} = \frac{\mu}{2(1+\alpha)}\left[(1-\alpha) \left( \lambda_s^2 + \lambda_\theta^2 + \lambda_s^{-2} \lambda_\theta^{-2} - 3 \right) + \alpha \left( \lambda_s^{-2} + \lambda_\theta^{-2} + \lambda_s^{2} \lambda_\theta^{2} - 3 \right) \right],
\end{equation}
where $\mu$ is the infinitesimal shear modulus of the elastic material and $0\leq \alpha \leq 1$ is a parameter that controls the deviation from a Hookean response. An important special case of \rr{mooney} is the neo-Hookean strain-energy function, which occurs when $\alpha=0$. The $(1+\alpha)$ in the denominator of \rr{mooney} is necessary to ensure that $\mu$ is the infinitesimal shear modulus of the material. \citet{fung1993} introduced a strain-energy function designed for the modelling of biological soft tissues such as healthy arterial tissue, which commonly feature strongly  strain-stiffening behaviour,
\begin{equation}
\hat{W} = \frac{\mu \left(e^{\Gamma \left(\lambda_s^2 + \lambda_\theta^2 + \lambda_s^{-1}\lambda_\theta^{-1} -3\right)} - 1\right)}{2 \Gamma} \label{Fung},
\end{equation}
where $\Gamma$ is a positive parameter representing the degree of strain stiffening. In the limit as $\Gamma$ approaches zero, \rr{Fung} also recovers the neo-Hookean strain-energy function.
%Generally the infinitesimal shear modulus of a three-dimensional strain-energy function may be found by evaluating, 
%\begin{equation}
%\left. \frac{\partial^2 \hat{W}}{\partial \lambda_s^2} \right|_{\lambda_s = \lambda_\theta =1} = 4 \mu,
%\end{equation}
%and we shall later use this value to non-dimensionalise any strain-energy function. We note that in \citet{yang1971} use the Mooney-Rivlin strain-energy function non-dimensionalise with respect to a constant which is not the  infinitesimal shear modulus.

As shown in Appendix A, the equilibrium equations in the tangential and normal directions are respectively given by,
\begin{subequations}\label{equib}
\begin{gather}
(\hat{r} \hat{T}_s)' -\hat{r}' \hat{T}_\theta =0, \label{equib1}\\
\hat{\kappa}_s \hat{T}_s + \hat{\kappa}_\theta \hat{T}_\theta = \hat{P},\label{equib2}
\end{gather}
\end{subequations}
where $\hat{P}(S)$ represents the pressure difference across the membrane in the normal direction, and the principal curvatures $\hat{\kappa}_s,\hat{\kappa}_\theta$ are defined in \rr{curvatures}. The system \rr{equib} contains terms involving $\hat{r}, \hat{r}',\hat{r}'',\hat{z}',\hat{z}''$, and is therefore a third order differential system, as $\hat{z}$ does not appear explicitly. Rearranging \rr{equib1} for $\hat{T}_\theta$ and using Codazzi's equation, \rr{codazzi2}, we may integrate \rr{equib2} to find,
\begin{equation}\label{intequib}
\hat{r}^2 \hat{\kappa}_\theta \hat{T}_s = \frac{\hat{F}}{2 \pi} + \frac{\hat{P} \hat{r}^2}{2} - \frac{1}{2} \int \dd{\hat{P}}{\hat{S}} \hat{r}^2 d \hat{S},
\end{equation}
which may be used, provided $\hat{P}$ is constant, to reduce the differential order of the system by one. We note that this integral exists for a membrane of any undeformed shape, and corresponds to the resultant in the $\hat{Z}$ direction. The integration constant $\hat{F}$ represents the total transverse force acting on the membrane \citep{nadler2006}.

\subsection{Wrinkling}
Using \rr{equib} we may calculate the deformed shape of the membrane, assuming that the stress resultants both stay positive over the region. If either stress resultant becomes negative then the theory as outlined above is not appropriate, as a true membrane can not support compressive stresses. The regions with such a negative compressive stress are assumed to have wrinkled, with the crests of the wrinkles perpendicular to the compressive stress direction; this is a local buckling instability \citep{jenkins1991}. We therefore introduce tension-field theory to consider a smoothed-out \lq pseudo-surface' that is in equilibrium, with the corresponding stress exactly zero everywhere. 

For the strain-energy functions considered here, the minimum of $W$ with respect to $\lambda_\theta$, subject to both stretches being positive, occurs  when $\lambda_\theta = \lambda_s^{-1/2} \equiv w(\lambda_s)$, with an equivalent expression for $\lambda_s$. This is termed the natural width of the strip in tension by \citet{pipkin1986}. Following \citet{pipkin1986} we introduce the tension-field theory by defining a relaxed strain-energy density function as,
\begin{equation}
W^{*}(\lambda_s,\lambda_\theta) =  
\left\{
	\begin{array}{lcc}
\hat{W}(\lambda_s,\lambda_\theta) & \lambda_s \geq 1, & \lambda_\theta \geq 1 \\
\hat{W}(w(\lambda_\theta),\lambda_\theta) & \lambda_s \leq w(\lambda_\theta), & \lambda_\theta>1 \\
\hat{W}(\lambda_s,w(\lambda_s)) & \lambda_s>1, & \lambda_\theta \leq w(\lambda_s) \\ 0  &\lambda_s<1, & \lambda_\theta<1
\end{array}
\right.
\end{equation}
which is the quasiconvexification of $\hat{W}$. Using this definition of $W^{*}$ we recover the normal behaviour in the taut region while resolving the wrinkled regions. When both stretches are less than unity, the membrane becomes slack, with no tension in either direction. If this occurs here then \rr{intequib} implies that $\hat{F}=0$ and thus there is no indentation, when $\hat{P}=0$.

%Instead of studying the detail of the wrinkled region, we consider a smoothed-out \lq pseudo-surface' that is in equilibrium, with $\hat{T}_\theta = 0$ everywhere \citep{libai1998}. The stretch in the meridonal direction on this pseudo-surface will not be the true stretch as wrinkles will have formed in the true surface, and we therefore introduce a pseudo-stretch, $\tilde{\lambda}_\theta$, which shall be defined in the same way as $\lambda_\theta$, $\tilde{\lambda}_\theta = \hat{r}/\hat{R}$. The true stretch $\lambda_\theta$ remains in the constitutive relations \rr{stressres}, but not in \rr{lambdas}, and we solve the additional constraint $\hat{T}_\theta=0$ to find $\lambda_\theta = \lambda_s^{-1/2}$. 

We find numerically that the wrinkled region, if it exists, occurs when $\hat{T}_\theta = 0$. We therefore have the following system of equations in the wrinkled region,
\begin{equation}\label{equibwrink}
\hat{\kappa}_s \hat{T}_s = \hat{P}, \;\; \left(\hat{r} \lambda_s^{1/2} \hat{W}_1(\lambda_s,\lambda_s^{-1/2})\right)'=0.
\end{equation}
When the pressure $\hat{P}$ is zero, \rr{equibwrink}$_1$ implies that either the membrane is slack (and thus there is no indentation) or $\kappa_s =0$. We may therefore integrate \rr{equibwrink} to give,
\begin{equation}\label{equibwrink2}
\frac{\hat{z}'}{\hat{r} \sqrt{\hat{r}'^2 + \hat{z}'^2}} = \text{const.}, \;\; \hat{r} \lambda_s^{1/2} \hat{W}_1(\lambda_s,\lambda_s^{-1/2}) = \text{const.},
\end{equation}
where the constants are determined by the values of $\hat{r}$ and $\hat{z}$ at the point at which the membrane first becomes wrinkled, $\hat{S}=\hat{S}_1$. 

The first order differential equation in \rr{equibwrink2} may then be integrated until $\tilde{\lambda}_\theta = \lambda_\theta$, at which point $\hat{T}_\theta$ given by \rr{stressres} becomes zero again, this is $\hat{S}=\hat{S}_2$. This process may then be repeated if the membrane became wrinkled again, if required. Clearly, in the pseudo-surface the lines in the $s$-direction must be straight as the associated curvature $\hat{\kappa}_s$ is zero. \citet{libai2002} present a general algorithm for dealing with axisymmetric wrinkling, and this approach conforms to their technique.

\citet{haughton1995} and \citet{haughton1996} state that the extent (in terms of $\hat{S}$) of the wrinkled region is determined solely by the tense theory, and suggest that if you use the tense theory even when $\hat{T}_\theta <0$, then $\hat{T}_\theta = 0$ at $\hat{S}_2$. However, this is contrary to what we find numerically here, and so we integrate the wrinkled region exactly.

\subsection{Indenter Profile}
We assume that the indenter is a cylinder with a curved tip, with a specified axisymmetric profile where the components in the $\mathbf{e}_r$ and $\mathbf{e}_z$ directions are given parametrically in terms of an angle $\phi$ from the negative $z$-axis. We therefore assume the indenter surface is described by 
\begin{gather}\label{indenter}
\psi(\phi) = \left\{ \begin{array}{lc}
 \hat{\rho} A(\phi) \mathbf{e}_r + \hat{\rho} B(\phi) \mathbf{e}_z, &  0 \leq \phi \leq \pi/2, \\
 \hat{\rho} \mathbf{e}_r + (\hat{\rho} B(\pi/2) + \hat{\rho} \phi)\mathbf{e}_z,  & \phi >\pi/2 
\end{array}
\right.
\end{gather}
where $\hat{\rho}$ is the radius of the cylindrical part of the indenter to which this curved tip is attached. We require that the axisymmetry around the $z$ axis is maintained and hence $A(0)=0$, and also that the indenter tip smoothly connects to the cylindrical part, $A(\pi/2) = 1, A'(\pi/2)=0$. In addition, we require the smoothness condition of $B'(0)=0$, which prevents the consideration of perfectly sharp pointed tips, although we may approach these. We shall also impose the restriction that the indenter profile is convex, although for some non-convex profiles replacing the profile by its convex hull may give the appropriate shape, although the calculation of the load exerted on the membrane may need to be adjusted.

If the indenter tip is spherical, then $A(\phi) = \sin \phi, B(\phi) = - \cos \phi$, where the negative sign merely ensures that we consider the  indentation to occur in the negative $\mathbf{e}_z$ direction. Spheroidal tips may be considered by setting $B(\phi) = - \gamma \cos \phi$, where $\gamma > 1$ for a prolate spheroid and $0<\gamma < 1$ for an oblate spheroid. As $\gamma \to 0$, the case of the indentation by a flat cylindrical punch is approached, while $\gamma \to \infty$ approximates indentation by a sharp tip. 

We note that it is possible to use an indenter which consists of just the \lq tip' by modifying \rr{indenter}, for instance when a solid sphere is allowed to deform the membrane under gravity as in the experiments of \citet{liu2001}. For a given applied force this may be modelled by a spherical-tipped cylindrical indenter, provided that the deformed surface does not touch the sides of the cylinder. Indeed, it has been stated by \citet{selvadurai2006} and \citet{nadler2006} that the deformed surface never makes contact with the cylindrical part prior to membrane penetration, although our preliminary experimental results and the second figure in \citet{nguyen2004mech} suggest otherwise, so perhaps that statement is a specification to their equations. In particular for the Varga strain-energy function we agree with \citet{nadler2006}, but not for general strain-energy functions. Hence we do allow the membrane to touch the cylindrical part of the indenter, where we set $\hat{r}(\hat{S}) \equiv \hat{\rho}$ and solve for $\hat{z}$, although these regions will be dashed on the load-indentation graphs presented in Section \ref{results}. %A simple generalisation of \rr{indenter} allows the theory to be modified for indenters without the cylindrical part.

\subsection{Solution Procedure}
We now non-dimensionalise with respect to the undeformed radius, $\hat{R}_L= \hat{R}(\hat{L})$, and the infinitesimal shear modulus, $\mu$,
\begin{gather}
(\hat{S},\hat{R},\hat{Z},\hat{H},\hat{r},\hat{z}, \hat{\rho},\hat{L}) = \hat{R}_L (S,R,Z,H,r,z,\rho,L), \;  (\hat{T}_s, \hat{T}_\theta) = \mu \nonumber
\hat{R}_L H (T_s,T_\theta), \\
\hat{P} = \mu P, \; \hat{F} =  \mu \hat{R}_L^2 H F, \; \hat{W} = \mu W.
\end{gather}
For flat membranes $\hat{R}_L \equiv \hat{L}$, but for a curved membrane this is not generally true. This rescaling leaves the previously stated equations for the stretches, stress resultants and curvatures unchanged, apart from dropping the hats from all variables and removing $\hat{H}$ from \rr{stressres}. The parameter $\rho$ is now a non-dimensional ratio relating the radius of the indenter to the radius of the undeformed membrane.

The problem may be divided into two distinct regions of $S$, that is the region in the undeformed geometry that is now in contact with the indenter, $0 \leq S \leq S_c$, and the outer region where the membrane is free, $S_c < S \leq L$. For a flat membrane, $S_c$ is the radius of the circle in the undeformed membrane which is in contact with the indenter in the deformed configuration, while for a curved reference surface $R(S_c)$ is the radius of the corresponding circle in the undeformed membrane. If wrinkling occurs in  the free region there will be four distinct regions instead, as we find numerically that the membrane returns to biaxial tension by $S=L$ in the examples considered here.

While the membrane is in contact with the indenter, we assume that the deformed shape follows the surface of the indenter and we therefore have,
\begin{equation}
r(S) = \rho A(\phi(S)), \; z(S) = - \delta + \rho B(\phi(S)),
\end{equation}
where $\delta$ is the depth of indentation to be found as part of the solution. Therefore in the contact region,
\begin{equation}\label{indentstretch}
\lambda_s = \rho \frac{\sqrt{\left(\dd{A}{\phi}\right)^2 + \left(\dd{B}{\phi}\right)^2}}{\sqrt{R'^2 + Z'^2}} \dd{\phi}{S}, \; \lambda_\theta = \rho \frac{A(\phi(S))}{R(S)},
\end{equation}
where we have used $\phi'(S) >0$. We may therefore evaluate the first equilibrium equation, \rr{equib1}, to find $\phi(S)$ given appropriate boundary conditions. The second equilibrium equation enables us to calculate the pressure $P$ exerted by the indenter on the membrane after calculating the membrane deformation, but is auxiliary to computing the deformation itself in the contact region.

\subsection{Boundary Conditions}
We now specify the boundary conditions in order to integrate the equations above. The requirement of axisymmetry implies that $\phi(0)=0$, while using the restrictions on $A$ and $B$ we find $\phi'(0) = \rho^{-1} \lambda_s(0)/A'(0)$, and we can therefore integrate \rr{equib1} to find $\theta(S)$ for a given stretch at the pole $\lambda_s(0)$.

In the region $S_c<S \leq L$, the membrane is governed by \rr{equib} with $P=0$, which gives us a third order system for $r$ and $z$, or $\lambda_s$ and $\lambda_\theta$. We require continuity of the membrane shape, and so we prescribe $r(S_c),r'(S_c),z'(S_c)$ to be those given from the inner region.
We allow for the membrane to be stretched prior to indentation, imposing a pre-stretch of $\lambda_p \geq 1$ on the material, that is $\lambda_p = r(L)$.
This pre-stretch corresponds to moving the circular boundary $S=L$ outwards in the $\mathbf{e}_r$ direction, inducing a deformation in the membrane; if the membrane is not initially flat then this will have the result of changing the shape of the initial profile by decreasing the vertical \lq height' of the membrane, as the material moves towards the level of the boundary circle.

Utilising a shooting method we specify the stretch at the pole, $\lambda_s(0)=\lambda_\theta(0) = \lambda_0$, and then guess a value of $S_c$. We then iterate on $S_c$ to satisfy the condition $\lambda_\theta(L) =  \lambda_p$. Having found the appropriate value of $S_c$ for a given $\lambda_0$ and $\lambda_p$, we may then calculate the value of the central displacement of the membrane which is given by,
\begin{equation}\label{zint}
z_h=\int_0^L \dd{z}{S} \; dS  = \int_0^{S_c} \dd{z}{S} \;dS + \int_{S_c}^L \dd{z}{S} \; dS,
\end{equation} 
where the solutions in the appropriate regions are substituted into each integral of \rr{zint}. From this we calculate the indentation depth, $\delta = z_h - Z_h$, where $Z_h = Z(L)-Z(0)$ is the original depth of the membrane. Having done this, we may vary $\lambda_0$ and repeat the process in order to solve for a variety of indentation depths, or equivalently for values of the applied force $F$.

%After using \rr{equib1} to calculate the indented profile, we may use \rr{equib2} to calculate the pressure applied in the normal direction, $p_n$. Having done this, we may use the normal vector,
%\begin{equation}
%\mathbf{n} = \frac{z' \mathbf{e}_r - r' \mathbf{e}_z}{\sqrt{r'^2+z'^2}},
%\end{equation}
%to find the total force applied by the indenter is given by,
%\begin{equation}
%F = \int_0^{S_c} p_n \frac{r'}{\sqrt{r'^2+z'^2}} dS.  
%\end{equation}

We therefore require the following parameters to be specified: the membrane prestretch $\lambda_p$, the radius of the cylindrical part of the indenter $\rho$, the undeformed membrane shape $(R,Z)$ and the shape of the indenter tip $(A,B)$. In addition, the strain-energy function must be specified, along with the value of any parameters included therein. The choice of strain-energy function is not a trivial matter in the large indentation regime, and we will show how the force required to indent the membrane is strongly dependent on the form of the strain-energy function used, past the initial linear regime.

\section{Results and Discussion}\label{results}
\begin{figure}[!htbf]
 \psfrag{aa}{$r$}
 \psfrag{bb}{$z$}
\centering \epsfig{figure=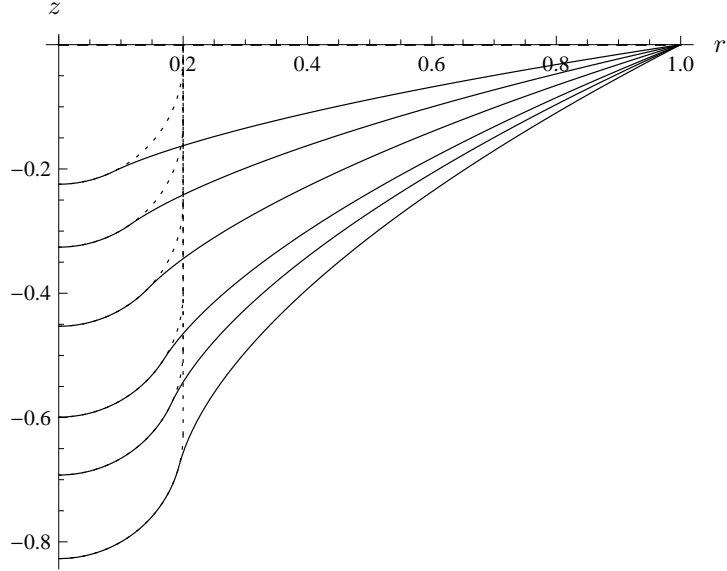,width=.6\textwidth}
\caption{Deformed configuration after indentation of a flat membrane by a spherical indenter for, from top to bottom, $\lambda_0 = 1.05,1.1,1.2,1.4,1.6,2$, with $\lambda_p = 1$ and $\rho = 0.2$. The strain-energy function used is the Mooney Rivlin with $\alpha=0.1$.}
\label{sphereindent}
\end{figure}
Figure \ref{sphereindent} shows how the deformed configuration changes with increasing $\lambda_0$ for a spherical indenter and a flat membrane, reproducing Figure 2 of \citet{yang1971}. It may be seen that the deformed shape goes from having positive curvature while in contact with the indenter to a negative curvature in the free region. This is as expected as equation \rr{equib2} implies that an odd number of the quantities $T_s,T_\theta,\kappa_s,\kappa_\theta$ must be negative in the free region, and therefore in unwrinkled parts of the membrane one of the curvatures must be negative. When the membrane is initially flat wrinkling is never observed, as found by \citet{haughton1995} for the fluid-loaded case.

%\begin{figure}[!htbf]
% \psfrag{aa}{$\delta$}
% \psfrag{bb}{$F$}
% \psfrag{lab1}{$\gamma = 0.01$}
%\psfrag{lab2}{$\gamma = 0.25$}
%\psfrag{lab3}{$\gamma = 1$}
%\psfrag{lab4}{$\gamma = 4$}
%\centering \epsfig{figure=indent2.eps,width=.6\textwidth}
%\caption{Load-indentation curves for a variety of spheroidal indenter tips with $A(\theta) = \sin \theta, B(\theta) = - \gamma \cos \theta$, otherwise same parameters as Figure \ref{sphereindent}.}
%\label{dispindent}
%\end{figure}

\begin{figure}[!htbf]
\psfrag{aa}{$\frac{\hat{\delta}}{\hat{R}_L}$}
\psfrag{bb}{$\frac{\hat{F}}{2 \pi \mu \hat{R}_L \hat{H}}$}
\centering \epsfig{figure=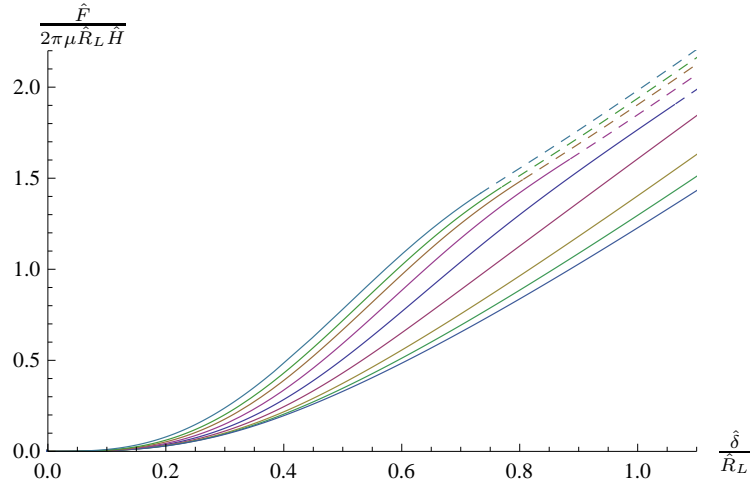,width=.6\textwidth}
\caption{
 a variety of spheroidal indenter tips with $A(\theta) = \sin \theta, B(\theta) = - \gamma \cos \theta, \lambda_p=1,\rho=0.2$. From top to bottom, $\gamma = 0.005,0.125,0.25,0.5,1,2,4,6,8,200$}
\label{varyshapecolour}\label{dispindent}
\end{figure}

%Having calculated the deformed shape of the membrane, we may use \rr{equib2} in order to calculate the pressure $P$ exerted by the indenter on the membrane. This enables us to calculate the total vertical load $w$ applied by the indenter for a given indentation. 

Having calculated the deformed shape of the membrane we may evaluate the transverse force $F$ in the free region, this corresponds to an indentation displacement $\delta$. Figure \ref{dispindent} shows a variety of non-dimensional force-indentation curves for spheroidal indenter tips of varying shape. As is intuitively obvious, blunter indenter tips require more force to achieve the same indentation depth. The two extremal curves in Figure \ref{dispindent} approach indentation with a flat punch and a point load respectively. In Figure \ref{dispindent}, and following figures, dashed lines represent where the membrane has come into contact with the cylindrical part of the indenter, which occurs sooner with flatter tips.
\begin{figure}[!htbf]
\psfrag{aa}{$\frac{\hat{\delta}}{\hat{R}_L}$}
\psfrag{bb}{$\frac{\hat{F}}{2 \pi \mu \hat{R}_L \hat{H}}$}
\psfrag{lab1}{\tiny neo-Hookean}
\psfrag{lab2}{\tiny Mooney-Rivlin, $\alpha=0.1$}
\psfrag{lab3}{\tiny Mooney-Rivlin, $\alpha=0.5$}
\psfrag{lab4}{\tiny Fung, $\Gamma = 1$}
\centering \epsfig{figure=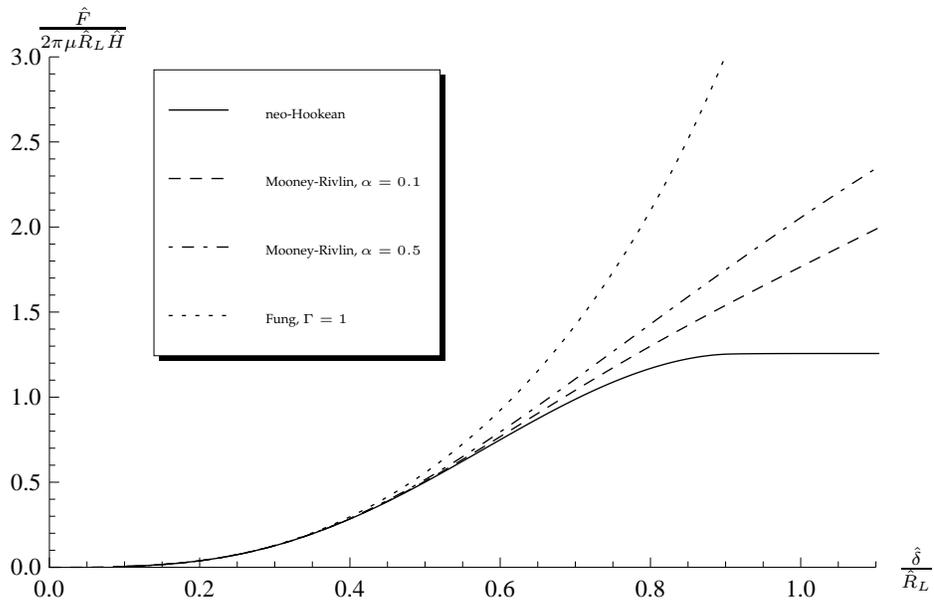,width=.8\textwidth}
\caption{Force-indentation curves for the indentation by a spherical indenter of a flat membrane modelled using the neo-Hookean, Mooney-Rivlin and Fung strain-energy functions with $\lambda_p=1,\rho=0.2$. }
\label{sevary}
\end{figure}
%\begin{figure}[!htbf]
%\psfrag{aa}{$\delta$}
%\psfrag{bb}{$F$}
%\psfrag{lab1}{\tiny NH}
%\psfrag{lab2}{\tiny MR, $\alpha=0.1$}
%\psfrag{lab3}{\tiny MR,  $\alpha=0.5$}
%\centering \epsfig{figure=indent3b.eps,width=.8\textwidth}
%\caption{Force-indentation curves for a variety of strain-energy functions, spherical indenter and a flat membrane with $\rho = 0.5$, $\lambda_p=1$}
%\label{sevary2}
%\end{figure}

Figure \ref{sevary} shows the load-indentation curves for a variety of strain-energy functions, where it may be seen that for this set of parameters up to $\delta \approx 0.4$ the material responses are identical, which is consistent to the results of \citet{pujara1978} regarding the inflation of circular membranes. Past this range the varied stress-strain behaviour of the strain-energy functions give different loads for the same indentation, and thereafter the specification of the strain-energy function becomes important. The Fung material model, generally used for biological tissues such as arteries and muscles, predicts much higher loads than the other models for the same amount of indentation, as expected due to the strain-stiffening behaviour the Fung model encapsulates. %Figure \ref{sevary2} shows similar behaviour for a larger indenter radius, $\rho = 0.5$, with the transition to non-linear behaviour now occurring at $\delta \approx 0.5$.

The neo-Hookean material model gives a horizontal asymptote in Figure \ref{sevary}, with a constant $F$ corresponding to an infinite range of indentation. This is a consequence of using the neo-Hookean strain-energy function outside its range of validity. %The Varga strain-energy function, used in \citet{nadler2006}, gives similar results. 
The Mooney-Rivlin material gives linear growth of $F$ with $\delta$, depending on the value of $\alpha$, while the Fung material shows an exponential growth.

In particular, when the membrane is touching the cylindrical portion of the indenter we have $r' = 0$, and therefore evaluating \rr{intequib} at $S=S_c$ gives,
\begin{equation}\label{Fcyl}
F = \left. 2 \pi R(S_c) \pafrac{W}{\lambda_s} \right|_{S=S_c},
\end{equation}
which explains how the response is dominated by the behaviour of the stress-strain relationship, as \rr{Fcyl} indicates that force is proportional to the stress. % We also note that as indentation increases the curvature of the force curves may become negative for the blunter tips. This appears to be associated with the choice of strain-energy function, and does not occur in the Fung strain energy function (with $\Gamma \geq 0.1$) for example. 

\begin{figure}[!htbf] 
\psfrag{aa}{$\frac{\hat{\delta}}{\hat{R}_L}$}
\psfrag{bb}{$\frac{\hat{F}}{2 \pi \mu \hat{R}_L \hat{H}}$}
\centering \epsfig{figure=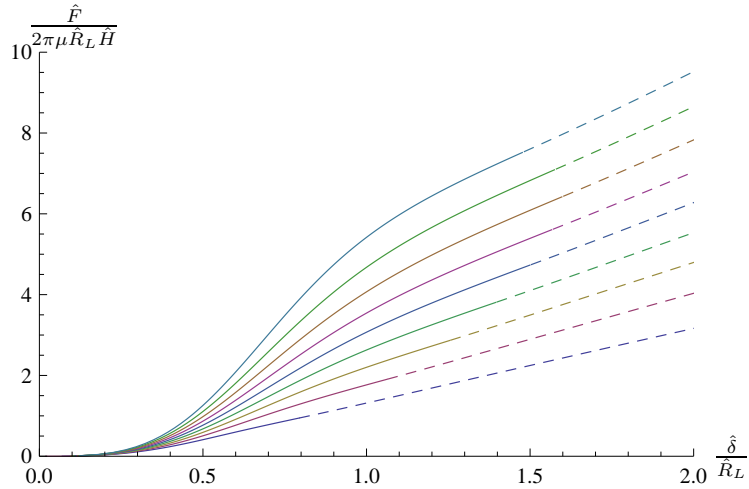,width=.6\textwidth}
\caption{Force-indentation curves for the indentation of a flat Mooney-Rivlin membrane by a spherical indenter of varying size, from bottom to top, $\rho=0.1,0.2,\ldots,0.9$}
\label{rhovary}
\end{figure}
Figure \ref{rhovary} shows how the size of the indenter affects the load behaviour, although the quantitative response behaviour is not affected. For large indentation depths the curves are parallel to each other, both before and after the membrane begins to touch the sides of the indenter. 

In all the force-indentation curves for flat membranes presented above, in the linear elastic limit the relation $F \propto \delta^3$ may be seen, corresponding to the Schwerin-type solution in linear elasticity \citep{komaragiri2005}. \citet{begley2004} derive an approximate solution with such a cubic relationship for a finite-sized spherical indenter and a flat membrane, but we do not agree numerically with their proportionality constant. 

\begin{figure}[!htbf] 
\psfrag{aa}{$\frac{\hat{\delta}}{\hat{R}_L}$}
\psfrag{bb}{$\frac{\hat{F}}{2 \pi \mu \hat{R}_L \hat{H}}$}\centering \epsfig{figure=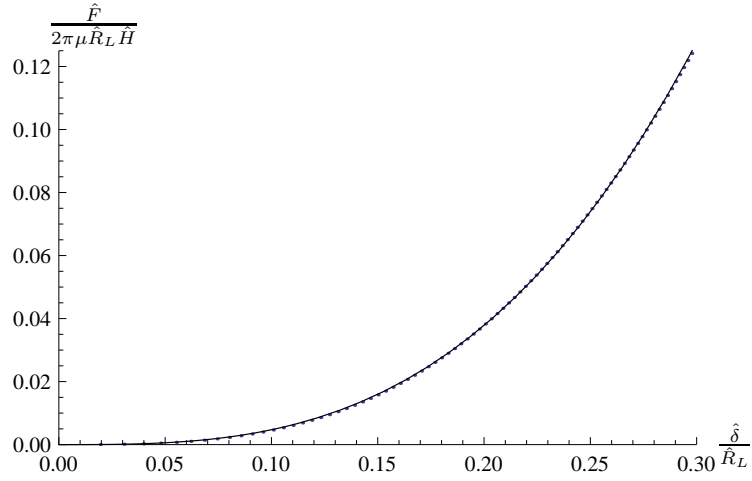,width=.6\textwidth}
\caption{Close-up of the force-indentation curve for a spherical indenter and a flat membrane with the Mooney Rivlin strain-energy function with $\alpha=0.1$, $\lambda_p=1$, $\rho=0.2$. Curve is given by $F=4.7292 \delta^3$.}
\label{sphcubic}
\end{figure}
 %Figure \ref{cubic} shows how, for the same selection of indenter tips in \ref{varyshapecolour}
%\citet{begley2004} derive an approximate solution for a finite-sized indenter, in which they find the following expression, rewritten in the notation here, for how the force and displacement are initially related,
%\begin{equation}
%\delta = (\frac{1}{\rho})^{1/12} (\frac{16 F}{9 \pi}^{1/3}
%\end{equation}
\begin{figure}[!htbf] 
\psfrag{aa}{$\frac{\hat{\delta}}{\hat{R}_L}$}
\psfrag{bb}{$\frac{\hat{F}}{2 \pi \mu \hat{R}_L \hat{H}}$}
\centering \epsfig{figure=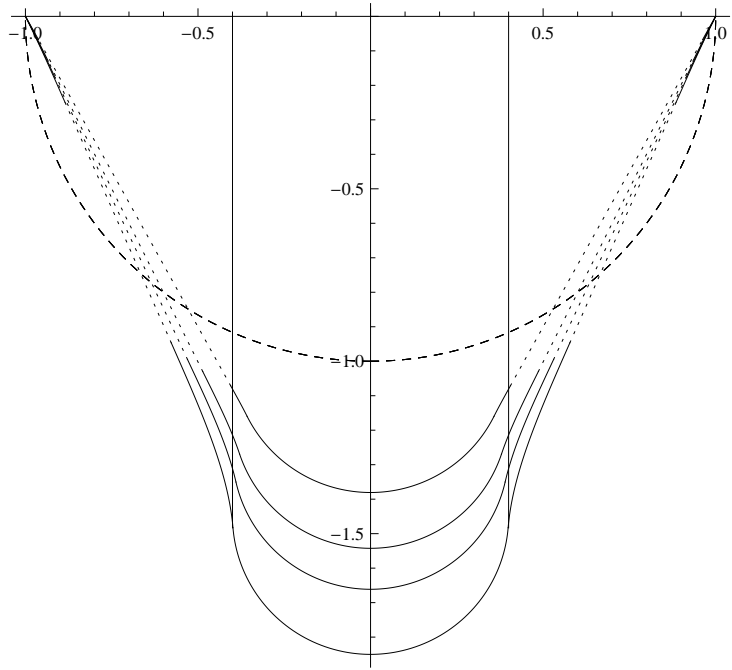,width=.6\textwidth}
\caption{Deformation of an initially spherical surface $R(S) = \sin S, Z(S) = - \cos S, \rho = 0.4, \lambda_p=1$ with a spherical indenter and a Mooney-Rivlin strain-energy function with $\alpha=0.1$. From top to bottom: $\lambda_0 = 1.2,1.4,1.6,2$. Dashed line shows the undeformed shape of the membrane, dotted sections of the deformed surface are wrinkled.}
\label{initspher}
\end{figure}

Figure \ref{initspher} shows the indentation of an initially spherical membrane by a spherical indenter of smaller radius. It can be seen that the deformed shape does not follow the curvature of the reference surface, which may be expected from the previous comment that $\kappa_s$ or $T_\theta$ must be negative, and from the fact that the governing equations do not include a term involving the reference curvatures, i.e. \rr{curvatures} but with $r$ and $z$ replaced by $R$ and $Z$. In order to include such a term in the governing equations, we would need to consider a shell theory where the energy depends on the curvature of the shell in addition to the stretching, as discussed in \citet{steigmann2001} for example. This increases the order of the governing equations from third to seventh order, and also introduces the question of determining how the strain-energy function depends on a further set of invariants; an additional two which depend only on the curvatures and three coupling terms between the curvatures and the stretches.

\begin{figure}[!htbf]
\psfrag{aa}{$\frac{\hat{\delta}}{\hat{R}_L}$}
\psfrag{bb}{$\frac{\hat{F}}{2 \pi \mu \hat{R}_L \hat{H}}$} \psfrag{lab1}{$\zeta = 0$}
\psfrag{lab2}{$\zeta = 0.125$}
\psfrag{lab3}{$\zeta = 0.25$}
\psfrag{lab4}{$\zeta = 0.5$}
\centering \epsfig{figure=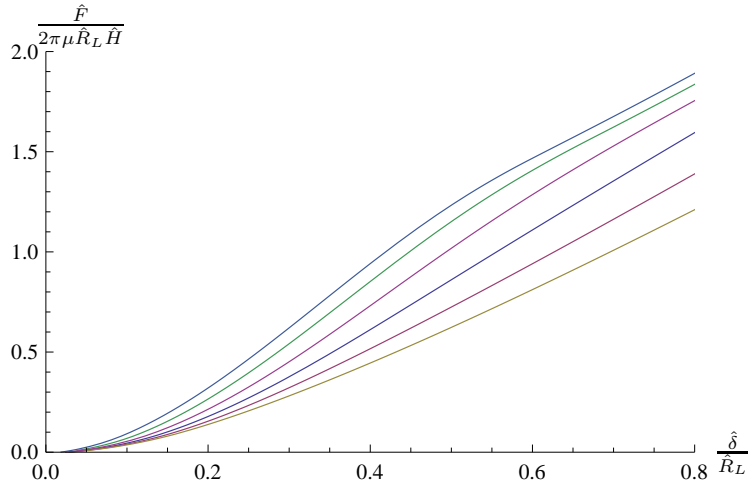,width=.6\textwidth}
\caption{Load-indentation curves for the indentation of an initially curved membrane with $R(S) = S, Z(S) = -\cos (\pi S/2)/4$ by various spheroidal indenter tips, from top to bottom: $\gamma = 0.25,0.5,1,2,4,8$ with $\lambda_p=1,\rho=0.2$}
\label{RZvary}
\end{figure}
\begin{figure}[!htbf]
\psfrag{aa}{$\frac{\hat{\delta}}{\hat{R}_L}$}
\psfrag{bb}{$\frac{\hat{F}}{2 \pi \mu \hat{R}_L \hat{H}}$} 
\centering \epsfig{figure=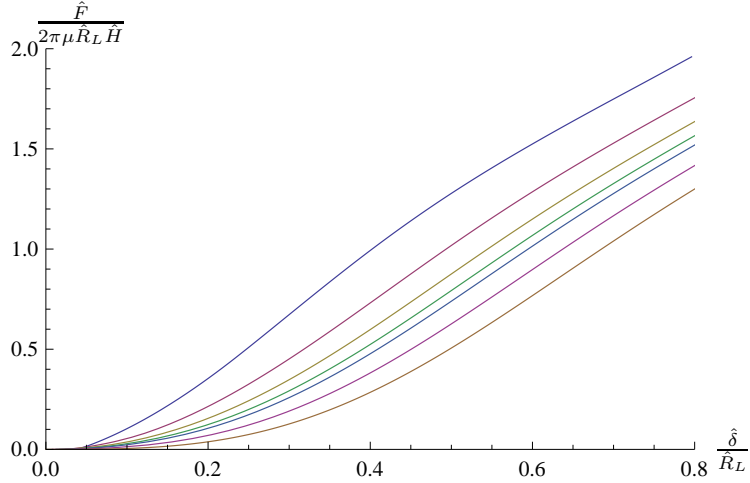,width=.6\textwidth}
\caption{Load-indentation curves for the indentation with a spherical indenter of an range of initially curved membranes with $R(S) = 2 S/ \pi, Z(S) = -\psi \cos S$, from top to bottom: $\psi=0.5,0.25,0.125,0.1,0.05,0$ and $\lambda_p=1,\rho=0.2$.}
\label{varyindent2}
\end{figure}

Figure \ref{varyindent2} shows the load-indentation curves for initially curved surfaces of the form, $R(S) = S, Z(S) = - \psi \cos ( \pi S /2), 0 \leq S \leq 1$. As $\psi \to 0$ we recover the flat membrane, and we see that as the reference surface departs from this shape the required load for greater indentation increases. 

For pronouncedly curved reference surfaces, $\psi \approx 0.5$, we find that the theory used here gives a vanishingly small load for a finite indentation displacement. As we take the numerical limit $\lambda_0 \to 1$, the calculated load tends towards zero but there is a non-zero displacement, $\delta \approx 0.09$ in the membrane shown in Figure \ref{varyindent2}. We believe that the reason this occurs is that the membrane wrinkles in the indentation region, where we are prescribing the displacement to be that of the indenter.

%For pronouncedly curved reference surfaces, $\psi \approx 0.5$, it is not possible to resolve the shape, and therefore the load, of the membrane for particularly small values of $\lambda_0$ using this method. We believe that the reason this occurs is that the membrane wrinkles in the indentation region, where we are prescribing the displacement to be that of the indenter. 

\begin{figure}[!htbf]
\psfrag{aa}{$\frac{\hat{\delta}}{\hat{R}_L}$}
\psfrag{bb}{$\frac{\hat{F}}{2 \pi \mu \hat{R}_L \hat{H}}$} 
\centering \epsfig{figure=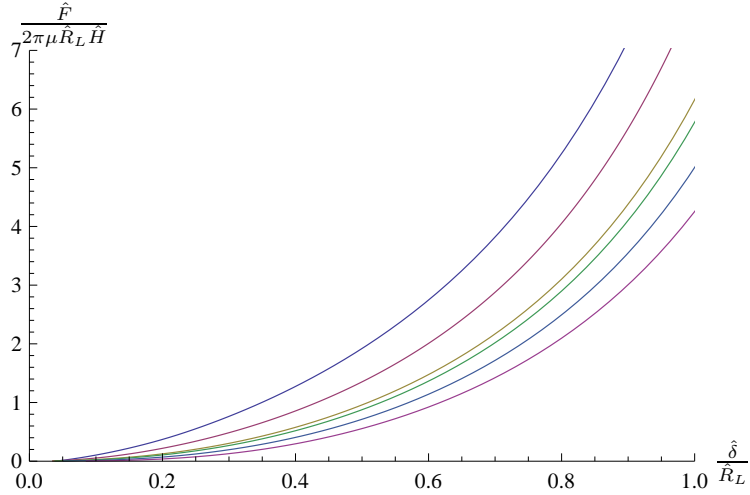,width=.6\textwidth}
\caption{Load-indentation curves for the indentation with a spherical indenter of an range of initially curved membranes with $R(S) = 2 S/ \pi, Z(S) = -\psi \cos S$, from top to bottom: $\psi=0.5,0.25,0.125,0.1,0.05,0$ and $\lambda_p=1,\rho=0.2$, using the Fung strain-energy function.}
\label{varyindent3}
\end{figure}

This limitation does not prevent us from solving for larger values of $\lambda_0$ due to the quasi-static solution procedure used here. Figure \ref{varyindent3} shows the same curved membrane as Figure \ref{varyindent2}, but with the Fung strain-energy function instead of the Mooney-Rivlin. The linear elastic behaviour of the two graphs is the same, but at large displacement or load the impact of the strain-energy function may be seen. 

%\begin{figure}[!htbf]
%\psfrag{aa}{$S$}
%\psfrag{bb}{}
%\psfrag{ee}{$r(S)$}
%\psfrag{ff}{$z(S)$}
%\psfrag{cc}{$T_s$}
%\psfrag{dd}{$T_\theta$}
%\centering \epsfig{figure=nonflatindent5.eps,width=.8\textwidth}
%\caption{a) Deformation of an initially spherical surface $R(S) = \sin S, Z(S) = - \cos S, \lambda_0=1.5,\rho = 0.2, \lambda_p=1$ with a spherical indenter and a NH strain-energy function. Dashed line shows the undeformed shape of the membrane. b) Stress resultants in the membrane, showing the region where $T_\theta$ becomes negative.}
%\label{curveddef2}
%\end{figure}
%
%Figure \ref{curveddef2} shows an initially spherical membrane cap indented by a smaller spherical indenter, although when the stress resultants are plotted it reveals that one of the principal stress resultants, the hoop stress $T_\theta$, becomes negative over a region of the membrane. Thus the actual deformed shape will incorporate wrinkles.  In order to properly describe this deformation it is necessary to either use a relaxed strain-energy function, or introduce the shell theory as discussed previously. 

When the membrane is flat, both the stress resultants stay positive and wrinkling does not occur. For curved membranes, it is possible to increase the radial pre-stretch $\lambda_p$ or the indentation distance $\delta$ to ensure that the membrane remains in tension in the deformed configuration. This finding agrees directly with that of \citet{haughton1996} for fluid filled membranes. For instance, if a spherical indenter nearly fills a spherical membrane, i.e. $\rho$ is close to one, then wrinkling is only found for very small levels of $\lambda_0$ with no pre-stretch, and eliminated entirely for slight amounts of pre-stretch, such as $\lambda_p \geq 1.1$. However, the imposition of the pre-strain will change the shape of the non-flat membrane prior to indentation, requiring knowledge of the shape of the membrane prior to the pre-strain, which may not be available in experimental situations.

\begin{figure}[!htbf]
\psfrag{aa}{$\delta$}
\psfrag{bb}{$S_c$}
\psfrag{lab1}{\tiny NH}
\psfrag{lab2}{\tiny MR, $\alpha=0.1$}
\psfrag{lab3}{\tiny MR, $\alpha=0.5$}
\psfrag{lab4}{\tiny Fung, $\Gamma = 1$}\centering \epsfig{figure=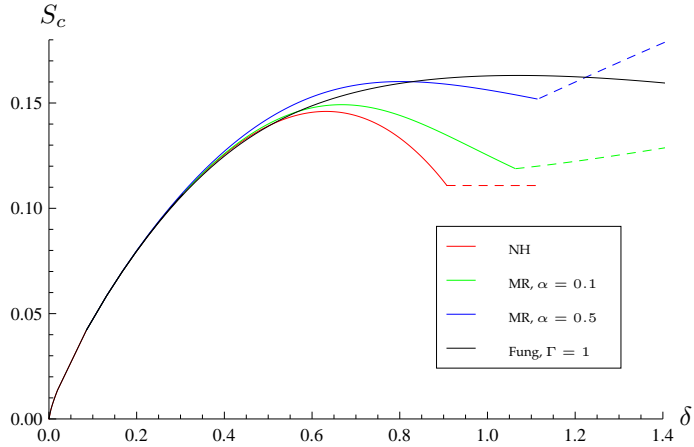,width=.6\textwidth}
\caption{Position of contact point $S_c$ as indentation depth changes for a indentation of a flat membrane with a spherical indenter, $\lambda_p=1, \rho=0.2$. The neo-Hookean (NH), Mooney-Rivlin (MR) and Fung strain-energy functions are shown.}
\label{scvary}
\end{figure}

As the indenter is pushed into a flat membrane the contact point $S_c$ initially increases as more of the indenter comes into contact with the material, before reaching a maximum value and then decreasing, as shown in Figure \ref{scvary}. This later decrease is due to the fact that, at larger indentation depth, a comparatively small area of the reference membrane is highly stretched over the indenter tip. Once the membrane begins to touch the cylindrical sides of the indenter the behaviour of $S_c$ is governed by the strain-energy function in the same manner as $F$. This behaviour is found by \citet{nadler2006} for the two-dimensional compressible Varga strain-energy function, although in this case the membrane never comes into contact with the cylindrical part of the indenter. 

\section{Conclusions}
We have shown how to extend the previously presented non-linear elasticity theories of indentation with a finite radius indenter to include the effects of different strain-energy functions, non-spherical indenter tips and initially curved membranes. The importance of  correctly specifying the strain-energy function is shown clearly, along with the effects that changing the shape of the indenter has. 

Indentation tests are a useful method of determining the properties of an elastic membrane at large strain, and with experimental data from a wide variety of shapes and sizes of indenter a suitable strain-energy function may be fitted to the data, giving more information than merely spherical indenter tips. One difficulty may be in identifying where the start of the force-indentation curve occurs experimentally, due to the detection limits of the equipment used, which may mean that the beginning of the curve is missed.

%The infinitesimal shear modulus may be estimated for experimental data from comparisons which only involve the linear regime, where the different strain-energy functions give the same response. 

As mentioned in the introduction, we plan to compare with experimental data from puncture force experiments on germinating seeds, with the aim of estimating the elastic properties of the endosperm using the theory presented here. The experiments will result in load-indentation curves, which we may then use to fit to the load-indentation curves resulting from the method here. In this case, the unknown parameters are the infinitesimal shear modulus and the strain-energy function of the material, and we shall attempt to characterize these through the  indirect indentation method.

\section*{Acknowledgements}
This work has been funded by an ERANET-PG grant for the vSEED project (BBSRC grant BBG02488X1). JRK additionally acknowledges the support of the 
Royal Society and the Wolfson Foundation.

\section*{Appendix}
\renewcommand{\theequation}{A.\arabic{equation}}
\setcounter{equation}{0}

In this appendix we shall provide a sketch of the derivation of the equilibrium equations given in \rr{equib}, using the membrane-like shell approach. For full details including proofs and error estimates of the derivation here see \citet{haughton2001} or \citet{haughton1978a}. Where possible we shall use upper case letters to describe the undeformed configuration and lower case letters to describe the deformed configuration. In addition, Greek indices vary over $(1,2)$, and the summation convention for repeated indices is used unless otherwise noted.

We consider a thin shell with position vector,
\begin{equation}
\mathbf{X} = \mathbf{Y}(\xi^1, \xi^2) + \xi^3 \mathbf{N}(\xi^1,\xi^2),
\end{equation}
where $(\xi^1,\xi^2)$ are curvilinear coordinates describing the midsurface of the shell, $-H/2 \leq \xi^3 \leq H/2$ is a thickness coordinate in the positive normal direction $\mathbf{N}$ and $H$ is the thickness of the undeformed shell. We assume that the shell is thin enough such that if $\eta$ is the minimum principal radii of curvature of the midsurface then,
\begin{equation}
\epsilon = \frac{H_{\textrm{max}}}{\eta} \ll 1,
\end{equation}
where $H_{\textrm{max}}$ is the maximum thickness throughout the shell.

We assume that the position vector $\mathbf{x}$ of the deformed membrane may be decomposed in the same way into a surface vector $\mathbf{y}$ and a thickness component in the deformed normal direction $\mathbf{n}$,
\begin{equation}
\mathbf{x} = \mathbf{y}(\zeta^1,\zeta^2) + \zeta^3 \mathbf{n}(\zeta^1,\zeta^2),
\end{equation}
where $(\zeta^1,\zeta^2)$ are coordinates describing the deformed surface of the shell and $-h/2 \leq \zeta^3 \leq h/2$ expresses the distance through the thickness direction of the membrane. \citet{haughton2001} show that, correct to within $\mathcal{O}(\epsilon)$, the surface $\zeta^3=0$ corresponds to the midsurface of the reference membrane $\xi^3=0$, so that $\mathbf{y}$ may be used as the midsurface of the deformed membrane. Expanding $\zeta^3$ around $\xi^3$, we may also find that $h= H \zeta^3_{,3}(\xi^1,\xi^2,0)$ to the same order. The tangent vectors to the undeformed and deformed surfaces may be written as,
\begin{equation}
\mathbf{G}_\alpha = \mathbf{Y}_{,\alpha}, \; \mathbf{g}_\alpha = \mathbf{y}_{,\alpha},
\end{equation}
and we also define dual vectors,
\begin{equation}
\mathbf{G}^\alpha = G^{\alpha \beta} \mathbf{G}_\beta, \; \mathbf{g}^\alpha = g^{\alpha \beta} \mathbf{g}_\beta,
\end{equation}
where $G^{\alpha \beta} G_{\alpha \beta} = \delta^{\alpha}_{\beta}, G_{\alpha \beta} = \mathbf{G}_\alpha \cdot \mathbf{G}_\beta, g^{\alpha \beta} g_{\alpha \beta} = \delta^{\alpha}_{\beta}, g_{\alpha \beta} = \mathbf{g}_\alpha \cdot \mathbf{g}_\beta$. The deformation gradient \citep{ogden2001} may be given by \citep{haughton2001},
\begin{equation}
\mathbf{F} = \pafrac{\mathbf{x}}{\mathbf{X}}=\pafrac{\mathbf{y}}{\mathbf{Y}} + \zeta^3_{,3}(\xi^1,\xi^2,0) \mathbf{n} \otimes \mathbf{N},
\end{equation}
from which we conclude that $\zeta^3_{,3}(\xi^1,\xi^2,0)$ is a principal stretch. We may write the two-dimensional deformation gradient as,
\begin{equation}\label{defgradx}
\pafrac{\mathbf{y}}{\mathbf{Y}} = \mathbf{g}_\alpha \otimes \mathbf{G}^{\alpha},
\end{equation}
which can then be directly computed. Following \citet{haughton2001} we now define an orthonormal set of basis vectors $(\mathbf{a_i})$ by,
\begin{equation}
\mathbf{a}_\alpha = \frac{\mathbf{g}_\alpha}{\left| \mathbf{g}_\alpha \right|} \;\;\;\; \textrm{no sum},
\end{equation}
which enables us to dispense with the distinction between covariant and contravariant tensors when we use $\mathbf{a}_\alpha$ as a basis \citep{fu2001}.
Referred to this basis, the general non-linear elasticity equilibrium equation $\textrm{div} \, \hat{\boldsigma} = \mathbf{0}$ becomes at zeroth order in $\epsilon$, \citep{haughton1978a, haughton2001}, 
\begin{subequations}\label{equilx}
\begin{gather}
\hat{\sigma}_{\mu \kappa, \mu} + \hat{\sigma}_{\mu \kappa} \mathbf{a}_{\mu,\nu} + \hat{\sigma}_{\mu \nu} \mathbf{a}_\kappa \cdot \mathbf{a}_{\mu, \nu} + \hat{h}_{, \mu} \hat{\sigma}_{\mu \kappa}/\hat{h} = 0, \quad \kappa=1,2 \label{equilx1}\\
\hat{\sigma}_{\mu \kappa} \mathbf{a}_3 \cdot \mathbf{a}_{\mu, \nu} + \hat{P}/\hat{h} =0\label{equilx2}
\end{gather}
\end{subequations}
where $\hat{\sigma}_{\alpha \beta}$ is the $(\alpha,\beta)$-th element of the stress tensor $\boldsigma$, $\hat{P}$ is a hydrostatic pressure applied to the inner surface and due to the use of the orthonormal basis vectors we have the additional definition $()_{,\mu} = \left| \pafrac{\mathbf{y}}{\zeta^\mu} \right|^{-1} \pafrac{()}{\xi^\mu}$. The third principal stress, $\hat{\sigma}_{33}$, is of order $\epsilon$ at most, and therefore does not appear in \rr{equibeqn}. This is often called the membrane assumption, $\hat{\sigma}_{33} = 0$.

It was shown by \citet{naghdi1977} that the equilibrium equations presented above may be shown to be equivalent to those derived by considering the two-dimensional surface theory, under certain assumptions on $\hat{P}$ and $\hat{\sigma}_{ij}$. This may be verified by a direct calculation of the equilibrium equations using the membrane theory detailed by \citet{steigmann1999} or \citet{steigmann2001} for instance. See \citet{haughton2001} for more details on the comparison between the two theories.

We now use the coordinate systems introduced in \rr{undefcon} and \rr{defcon} to describe the undeformed and deformed configurations respectively. We may therefore associate the two curvilinear coordinate systems $(\xi^\alpha),(\zeta^\alpha)$ with the arclength and polar angle,
\begin{equation}
\xi^1 = A \Theta, \xi^2 = S, \;\; \zeta^1 = A \theta, \zeta^2 = S,
\end{equation}
where $A$ is a representative length. The orthonormal basis then becomes,
\begin{equation}\label{adef}
\mathbf{a}_1 = \mathbf{e}_\theta, \; \mathbf{a}_2 = \frac{\hat{r}' \mathbf{e}_r + \hat{z}' \mathbf{e}_z}{(\hat{r}'^2+\hat{z}'^2)^{1/2}}, \; \mathbf{a}_3 = \mathbf{a}_1 \wedge \mathbf{a}_2  = \frac{\hat{z}' \mathbf{e}_r - \hat{r}' \mathbf{e}_z}{(\hat{r}'^2+\hat{z}'^2)^{1/2}},
\end{equation}
where a prime represents differentiation with respect to $\hat{S}$. It may be seen that these vectors are unit vectors in the meridional, tangential and normal directions of the deformed surface. A similar orthonormal basis $\mathbf{A}_\alpha$ may be described with respect for the undeformed configuration with upper case letters replacing all equivalent lower case ones. The deformation gradient \rr{defgradx} is then given by,
\begin{align}
\mathbf{F} &= \frac{\hat{r}}{\hat{R}} \mathbf{e}_\theta \otimes \mathbf{e}_\theta + \left(\hat{r}' \mathbf{e}_r + \hat{z}' \mathbf{e}_z \right) \otimes \left(\frac{\hat{R}' \mathbf{e}_r + \hat{Z}' \mathbf{e}_z}{\hat{R}'^2 + \hat{Z}'^2}\right) + \frac{\hat{h}}{\hat{H}} \mathbf{n} \otimes \mathbf{N} \nonumber \\
&=\frac{\hat{r}}{\hat{R}} \mathbf{e}_\theta \otimes \mathbf{e}_\theta+
\frac{\sqrt{\hat{r}'^2+\hat{z}'^2}}{\sqrt{\hat{R}'^2+\hat{Z}'^2}} \left(\frac{\hat{r}' \mathbf{e}_r + \hat{z}' \mathbf{e}_z}{\sqrt{\hat{r}'^2+\hat{z}'^2}}\right) \otimes \left(\frac{\hat{R}' \mathbf{e}_r + \hat{Z}' \mathbf{e}_z}{\sqrt{\hat{R}'^2+\hat{Z}'^2}}\right) + \frac{\hat{h}}{\hat{H}} \mathbf{n} \otimes \mathbf{N}.\label{defff}
\end{align}
which shows that the principal directions are the meridional, tangential and normal directions respectively. As the deformation gradient can be written in a diagonal form, the stress tensor must also be diagonal with respect to the same basis, and so the only non-zero terms of $\hat{\boldsigma}$ relative to this basis are $\hat{\sigma}_{11}$ and $\hat{\sigma}_{22}$. Additionally, using this basis we may immediately see the principal stretches in the three directions as the coefficients in \rr{defff}.
Using \rr{adef} in \rr{equilx}, along with some rearrangement and noting that $()_{,2}$ in \rr{equilx} becomes $()'/\sqrt{\hat{r}'^2 + \hat{z}'^2}$, we find the equilibrium equations, 
\begin{gather}\label{equibeqn}
\pafrac{(\hat{h} \hat{r} \hat{\sigma}_{11})}{\theta} =0, \\
\pafrac{(\hat{h} \hat{r} \hat{\sigma}_{22})}{S} =  \hat{h} \hat{r}' \hat{\sigma}_{11},\\
\hat{\sigma}_{11} \frac{\hat{z}'}{r \sqrt{\hat{r}'^2+\hat{z}'^2}} + \hat{\sigma}_{22} \frac{\hat{z}'' \hat{r}' - \hat{r}'' \hat{z}'}{(\hat{r}'^2+\hat{z}'^2)^{3/2}} - \frac{\hat{P}}{\hat{h}} = 0.
\end{gather}
The first equation in \rr{equibeqn} is immediately satisfied due to the restriction of axisymmetry in the deformed configuration. The remaining two equations are the governing equations for the deformation, and represent force balances in the tangential and normal directions respectively. In \rr{equibeqn}$_3$ the coefficients of the stresses are actually the principal curvatures in the deformed configuration, 
\begin{equation}\label{curvatures}
\hat{\kappa}_s =  \frac{1}{\hat{r}'}\left(\frac{\hat{z}'}{\sqrt{\hat{r}'^2+\hat{z}'^2}}\right)' = \frac{-1}{\hat{z}'}\left(\frac{\hat{r}'}{\sqrt{\hat{r}'^2+\hat{z}'^2}}\right)' , \; \hat{\kappa}_\theta = \frac{\hat{z}'}{\hat{r} \sqrt{\hat{r}'^2+\hat{z}'^2}},
\end{equation}
where $\kappa_s, \kappa_\theta$ are the principal curvatures in the circumferential and radial directions respectively. These curvatures may be found from the components of the embedded deformed curvature tensor $\boldkappa = \kappa_{\alpha \beta} \mathbf{G}^{\alpha} \otimes \mathbf{G}^{\beta}$ using the definition $\kappa_{\alpha \beta} = \mathbf{n} \cdot \mathbf{g}_{\alpha, \beta}$, and then referred to the appropriate basis by $\kappa_\theta = \kappa_{11} g^{11}, \kappa_s = \kappa_{22} g^{22}$.

When the undeformed surface is flat or spherical we have $\sqrt{\hat{R}'^2 + \hat{Z}'^2} = 1$ and the radicals in equation \rr{curvatures} may be expressed neatly including the principal stretch $\lambda_s$, but for a general surface this is not the case. We also note that the curvatures are always related by Codazzi's equation,
\begin{equation}\label{codazzi2}
(\hat{r} \hat{\kappa}_\theta)' = \hat{r}' \hat{\kappa}_s. 
\end{equation}
which enables us to integrate one of the equilibrium equations under certain conditions, see \rr{intequib}.

\clearpage
\bibliographystyle{chicago}
\bibliography{./Indentref,./References}

\end{document}